\documentclass[aps,prl,groupedaddress,showpacs]{revtex4}
\usepackage{graphicx}
\usepackage{dcolumn}
\usepackage{bm}
\usepackage{color}

\begin{document}

\title{Bose-Einstein condensation in quasi-2D systems: \\
applications to high T$_c$ superconductivity}
\author{C. Villarreal$^1$ and M. de Llano$^2$ }
\date{\today }
\affiliation{$^{1}$Instituto de F\'{\i}sica, Universidad Nacional Aut\'{o}noma de M\'{e}%
xico, Circuito de la Investigaci\'{o}n Cient\'{\i}fica, Ciudad
Universitaria, CP 04510 Distrito Federal, M\'{e}xico \\
$^{2}$Physics Department, University of Connecticut, Storrs, CT 06269 USA}
\address{$^{\ast }$Instituto de Investigaciones en Materiales, Universidad
Nacional Aut\'{o}noma de M\'{e}xico\\
Apdo. Postal 70-360, 04510 M\'{e}xico, DF, MEXICO}

\begin{abstract}
We describe high-$T_{c}$ superconductivity in layered materials
within a BCS theory as a BEC of massless-like Cooper pairons
satisfying a linear dispersion relation and propagating within
quasi-2D layers of finite width $\delta $ defined by the charge
distribution about the CuO$_{2}$ planes. We obtain a closed formula
for the critical temperature $T_{c}\propto \sqrt{\delta \vert
\epsilon_0 \vert} /\lambda_{ab} $, where $\epsilon_{0}$ is the
binding energy of Cooper's pairs, and $\lambda_{ab}$ the average
in-plane penetration depth. This formula reasonably reproduces
empirical values of superconducting transition temperatures for
several different cuprate materials near the optimal doping regime,
as well as for YBCO films with different doping degrees.

$^{\ast }$Permanent address.
\end{abstract}

\maketitle

\address{and\\
*Instituto de Investigaciones en Materiales, Universidad Nacional Aut\'{o}noma de M\'{e}xico, Apdo. Postal 70-360, 04510 M\'{e}xico, DF, MEXICO}

%\eads{ \mailto{CV <carlos@fisica.unam.mx>},
%\mailto{MdL<mdellano@fis.unam.mx>}}

%\section{Introduction.}

Recent experimental studies of the spectral intensity of photoelectron
emission in high $T_{c}$ cuprate superconductors have provided evidence that
bound electron Cooper pairs (pairons) already exist at temperatures higher
than the critical transition temperature \cite{yang08}. This finding is
consistent with several theoretical proposals that suggest that high $T_{c}$
superconductivity (HTSC) originates from a 2D Bose-Einstein condensate (BEC)
of pairons pre-existing above $T_{c}$, coupled through a BCS-like phonon
mechanism \cite{fujita96,dellano98}. The 2D character of the phase
transition is associated with the layered structure of cuprates, which in
the case of YBaCu$_{3}$O$_{7-y}$ (YBCO) consists of a succession of layers
along the c-axis with a unit cell of length $c\approx 12\mathring{A}$, and
the chemical composition CuO-BaO-CuO$_{2}$-Y-CuO$_{2}$-BaO-CuO. It is widely
accepted that the CuO$_{2}$ planes, which in the case of YBCO are
equidistant from the central Y atom by a distance $\approx 1.5\mathring{A}$,
are mainly responsible for the superconductivity in cuprates. Contour plots
of the charge distribution derived from energy-band-structure calculations
for YBCO reveal \cite{krakauer88} that the SC charge carriers are mainly
concentrated within a shell of width $\delta \approx 2.15\mathring{A}$ about
the CuO$_{2}$ plane. We may thus assume that the number of SC charges per
unit area, $n^{2D}$,\ is approximately determined by $n^{2D}=\delta n^{3D}$,
where $n^{3D}$ is the volume charge density.

Following closely the formalism developed by Cooper \cite{cooper56} let us
consider a quasi-2D system of fermions with effective mass $m^{\ast }$,
kinetic energies $\varepsilon _{k_{1}}=\hbar ^{2}k_{1}^{2}/2m^{\ast }$ and $%
\varepsilon _{k_{2}}=\hbar ^{2}k_{2}^{2}/2m^{\ast }$, relative wave number $%
\mathbf{k}=%
%TCIMACRO{\U{bd}}%
%BeginExpansion
{\frac12}%
%EndExpansion
(\mathbf{k}_{1}-\mathbf{{k}_{2}}$), and center of mass (CM) wave vector $%
\mathbf{K}=\mathbf{k}_{1}+\mathbf{k}_{2}$. The fermions interact pairwise
via the Cooper model interaction
\begin{equation}
V_{kk^{\prime }}=-V_{0}\ \ \mathrm{if}\ \ k_{F}<|\mathbf{k}\pm
%TCIMACRO{\U{bd}}%
%BeginExpansion
{\frac12}%
%EndExpansion
\mathbf{K}|,|\mathbf{k}^{\prime }\pm
%TCIMACRO{\U{bd}}%
%BeginExpansion
{\frac12}%
%EndExpansion
\mathbf{K}|<K_{max},  \label{cooper}
\end{equation}%
and $V_{k,k^{\prime }}=0$, otherwise, with $V_{0}>0$. Here, $K_{max}=2\sqrt{%
k_{F}^{2}+k_{D}^{2}}$, where $k_{F}$ and $k_{D}$ are defined, respectively,
by the Fermi energy $E_{F}\equiv \hbar ^{2}k_{F}^{2}/2m^{\ast }$, and by the
Debye frequency $\omega _{D}$\ of lattice phonons of maximum energy $\hbar
\omega _{D}\equiv \hbar ^{2}k_{D}^{2}/2m^{\ast }$. A simple geometric
construction shows that bound pairs will form only if the tip of vector $%
\mathbf{k}$ lies within the overlap of the two rings defined by condition (%
\ref{cooper}) in k-space \cite{dellano07}. Thus, fermions lying outside this
overlapping region are unpairable.

The variation of the bound state energy with the CM momentum $\hbar \mathbf{K%
}$ was discussed by Schrieffer \cite{schrieffer64} by considering the Cooper
condition for the energy eigenvalues of the fermion pair:
\begin{equation}
V_{0}\sum_{\mathbf{k}}\frac{1}{\hbar ^{2}k^{2}/m^{\ast }+\hbar
^{2}K^{2}/4m^{\ast }-2E_{F}-\varepsilon _{K}}=1  \label{condition}
\end{equation}%
where the summation is restricted to values of $\mathbf{k}$ allowed by
interaction (\ref{cooper}). The summation may be approximated by an
integration over the density of states, $N(\epsilon )$, which in 2D is
independent of the energy: $N(\epsilon )\equiv N_{0}=m/2\pi \hbar ^{2}$. In
the small $K$ limit, it is found that
\begin{equation}
\varepsilon _{K}=\varepsilon _{0}+c_{1}\hbar K+O(K^{2})  \label{dispersion}
\end{equation}%
where $\varepsilon _{0}$ is the pairon binding energy \cite{cooper56} at
zero momentum, namely%
\begin{equation}
\varepsilon _{0}=-\frac{2\hbar \omega _{D}}{e^{2/V_{0}N_{0}}-1}
\label{epsilon0}
\end{equation}%
and $c_{1}=2v_{F}/\pi $ with $v_{F}$ the Fermi velocity. Thus, Eq.(\ref%
{dispersion}) provides an approximate dispersion relation, linear at leading
order, rather than quadratic. As a consequence, all \textit{excited} pairons
behave like free massless particles with a common group velocity $%
c_{1}=\hbar ^{-1}d\varepsilon _{k}/dk$, but a variable energy determined by
their CM momenta $\hbar K$. As pointed out by Schrieffer \cite{schrieffer64}%
, the linear dispersion relation implies that in order for a pairon to
remain bound ($\varepsilon _{K}<0$) its maximum allowed CM wavenumber is $%
K_{0}=|\varepsilon _{0}|/c_{1}$, since pairons with $K>K_{0}$ will break up.

These features are not exclusive of the Cooper model interaction (\ref%
{cooper}). For example, an attractive inter-fermion delta potential in 2D,
imagined regularized to support a single bound state of energy $-B_{2}$,
leads \cite{dellano07} to the dispersion relation $\varepsilon
_{K}=\varepsilon _{0}+c_{1}\hbar K+\left[ 1-(2-16/\pi ^{2})E_{F}/B_{2}\right]
\hbar ^{2}K^{2}/4m+O(K^{3})$, with a \textit{linear} leading term and $%
c_{1}=2v_{F}/\pi $ as with the Cooper model interaction. It is noteworthy
that only in the vacuum limit $v_{F}\rightarrow 0\Rightarrow
E_{F}\rightarrow 0$ does this latter dispersion relation lead to the
expected \textit{quadratic} form $\varepsilon _{K}=\varepsilon _{0}+\hbar
^{2}K^{2}/4m$. Then, in either case, the linear term is a consequence of the
presence of the Fermi sea. It has been reported that a linear dispersion
relation leads to very good fits of the BEC condensate-fraction curves for
quasi-2D cuprates \cite{dellano06} as well as for 3D and even 1D
superconductors.

Accordingly, we describe the total amount of charge carriers by means of an
ideal mixture of non-interacting unpaired fermions, and breakable pairons
with a linear dispersion relation \cite{dellano98,casas01}. The fermion
number per unit area is $n_{f}=n_{f1}+n_{f2}$, where $n_{f1}$ and $n_{f2}$
denote the number densities of unpairable, and pairable fermions,
respectively. Unpairable fermions obey a usual Fermi-Dirac distribution,
while the pairable fermion density at an arbitrary temperature $T$ can be
calculated as:
\begin{equation}
n_{f2}(T)=2\ \big[\ n_{0}^{2D}(T)+n_{0<K\leq K_{0}}^{2D}(T)\ \big]%
+n_{f2}^{u}(T),  \label{fermi}
\end{equation}%
where $n_{0}^{2D}$ represents the bosonic density of Cooper pairs with CM
wave vector $K=0$, $n_{0<K<K_{0}}^{2D}$ the equivalent quantity with $%
0<K<K_{0}$, and $n_{f2}^{u}$, the number density of \textit{pairable but
unpaired} fermions. By asserting that in thermal equilibrium this kind of
fermions arises precisely from broken pairons \cite{dellano98}, we identify $%
n_{f2}^{u}(T)=2n_{K_{0}<K<K_{max}}^{2D}(T)$. On the other hand, at $T=0$ all
pairable fermions should belong to the condensate (although, according to
\cite{casas01}, this is strictly valid only in the strong coupling limit),
and then $n_{f2}(0)=2n_{0}^{2D}(0)\equiv 2n^{2D}$, where $n^{2D}$ is the
total boson number per unit area. Summarizing, the number equation for
pairable fermions may be re-expressed in terms of bosonic quantities only: $%
n^{2D}=n_{0}^{2D}(T)+n_{0<K\leq K_{0}}^{2D}(T)+n_{K_{0}<K\leq
K_{max}}^{2D}(T)$. The last two terms in this equation belong to adjacent
momental regions, and so they may be merged into a single 2D integral of a
Bose-Einstein distribution:
\begin{equation}
n^{2D}=n_{0}^{2D}(T)+\frac{1}{(2\pi )^{2}}\int_{0}^{K_{max}}\frac{d^{2}K}{%
z^{-1}e^{\beta \varepsilon _{K}}-1},  \label{bose}
\end{equation}%
where $\beta \equiv 1/k_{B}T$, $z\equiv e^{\beta \mu }$ is the fugacity and $%
\mu $ the chemical potential. When the energy-shifted dispersion relation $%
\varepsilon _{K}=\hbar c_{1}K$ is introduced in (\ref{bose}) the integral
may be evaluated by changing to the variable $x\equiv \beta \hbar c_{1}K$.
Taking into account that $c_{1}\approx v_{F}$ and $K_{max}\approx
2K_{F}(1+K_{D}^{2}/2K_{F}^{2})$ the upper integration limit in (\ref{bose})
must be very large, namely $x_{max}=\beta \hbar v_{F}k_{F}\approx
E_{F}/k_{B}T\gg 1$. The last inequality is consistent with the maximum
empirical value for the ratio $k_{B}T_{c}/E_{F}\leq 0.05$ reported \cite%
{uemura04}\ in exotic SCs, including cuprate SCs. Given the rapid
convergence of Bose integrals, the upper integration limit $x_{max}$\ may
safely be taken as infinite in (\ref{bose}), and the integral can be
evaluated exactly by expanding the integrand in powers of $ze^{-x}$. This
gives%
\begin{equation}
n^{2D}=n_{0}^{2D}(T)+\frac{(k_{B}T)^{2}}{\pi \hbar ^{2}c_{1}^{2}}%
\sum_{n=1}^{\infty }\frac{z^{m}}{n^{2}}.  \label{density}
\end{equation}%
The critical BEC temperature $T_{c}$ is now determined by solving (\ref%
{density}) for $n_{0}^{2D}(T_{c})=0$ and $z(T_{c})=1$. We obtain
\begin{equation}
T_{c}=\frac{\hbar c_{1}}{k_{B}}\left( \frac{2\pi n^{2D}}{\zeta (2)}\right)
^{1/2}  \label{critical1}
\end{equation}%
where the Riemann Zeta function of order two $\zeta (2)=\pi ^{2}/6$.

\begin{figure}[t]
\begin{center}
\includegraphics[scale=0.5]{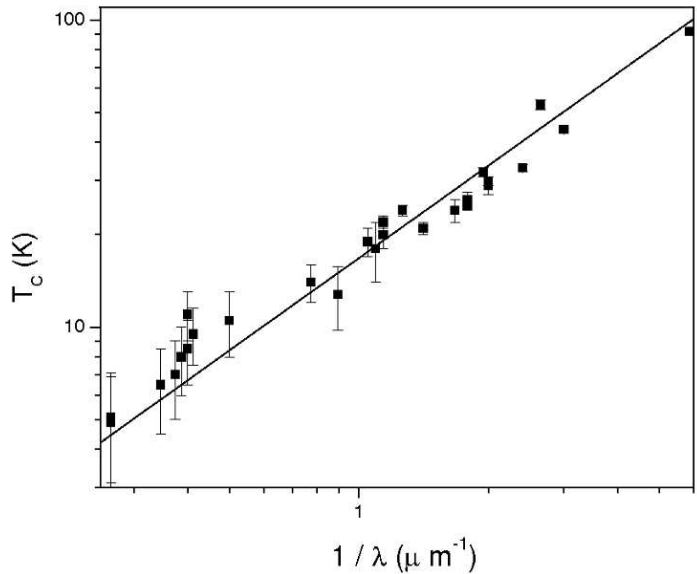}
\end{center}
\caption{Comparison of experimental critical temperatures vs. theoretical
predictions as given by Eq.(\protect\ref{critical3}) as a function of the
zero-temperature (inverse) penetration length $\protect\lambda_{ab}^{-1}$
for YBCO systems with different doping degrees.}
\end{figure}

A crucial element in the evaluation of (\ref{critical1}) is to reliably
estimate the fraction of charge carriers that actually contribute to the
supercurrent. The 3D charge carrier density $n^{3D}$ is usually determined
from measurements of London penetration depths $\lambda $. This parameter
gives an estimate of the current that causes partial rejection of an applied
external magnetic field in the superconductor. Within the framework of this
model, the supercurrent $\mathbf{J}_{s}$ is due to massless-like pairons
with a charge $2e$, density $n^{3D}=n_{s}/2$ (with $n_{s}$ the superfluid
density), and moving with a speed $c_{1}$. In that case, $\mathbf{J}%
_{s}=(2e)n^{3D}c_{1}\hat{\mathbf{K}}$, where $\hat{\mathbf{K}}\equiv \mathbf{%
K}/K$ \cite{fujita96}. In order to obtain the expression for the associated
penetration depth, we follow similar steps as those discussed in (\cite%
{fujita96}) and consider the contour integral of the pairon wavefunction
phase in a homogeneous medium, and in presence of an external magnetic field
$\mathbf{B}=\nabla \times \mathbf{A}$. The integral along any closed path
vanishes:
\begin{equation}
\oint \left( \hbar \mathbf{K}+\frac{2e}{c}\mathbf{A}\right) \cdot d\mathbf{r}%
=0.  \label{path}
\end{equation}%
Here, $c$ is the speed of light in vacuum. By eliminating $\mathbf{K}$ in
terms of $\mathbf{J}_{s}$, and using Stoke's theorem to evaluate (\ref{path}%
), we get a modified version of the London equation:
$\mathbf{J}=-\Lambda _{p}\mathbf{A}$, where $\Lambda _{p}\equiv
4e^{2}c_{1}n^{3D}/\hbar cK$. By taking now the curl of the modified
London equation and introducing Ampere law $\nabla \times
\mathbf{B}=(4\pi /c)\mathbf{J}_{s}$, it follows that the magnetic
field satisfies $\nabla ^{2}\mathbf{B}=\lambda ^{-2}\mathbf{B}$,
where
\begin{equation}
\frac{1}{\lambda ^{2}}\equiv \frac{(2e)^{2}}{c^{2}}\left( \frac{4\pi
c_{1}n^{3D}}{\hbar K}\right) .  \label{london}
\end{equation}%
This implies that the magnetic field decays exponentially within a
distance $\lambda $ measured from the superconductor-vacuum
interface. This expression is equivalent to the standard London
formula $\lambda _{L}^{-2}=4\pi e^{2}n_{s}/m^{\ast }c^{2}$ for
massive superelectrons with momentum $\hbar K=m^{\ast }c_{1}$.
Notice that, depending on the wave number $K$, $\lambda $ varies
between its minimum value $\lambda =0$ for $K=0$, corresponding to
perfect diamagnetism, and its maximum, say $\lambda _{0}$, for
$K=K_{0}$, which corresponds to pairon breakup. It seems natural to
assume that this latter condition determines the experimentally
observed value of the in-plane penetration
depth at zero temperature, namely $\lambda _{0}=\lambda _{ab}(T=0)$. Here, $%
\lambda _{ab}^{-1}(0)=\lambda _{a}^{-1}(0)+\lambda _{b}^{-1}(0)$ is the
geometric mean of the values of this parameter measured along the
crystallographic directions $a$ and $b$. We introduce now the relation $%
n^{2D}=\delta n^{3D}$ and employ the dispersion relation (\ref{dispersion})
to eliminate $K_{0}$ from $\lambda _{0}$. The final expression for the 2D
density of superconducting charge carriers is
\begin{equation}
n^{2D}=\frac{e^{2}}{c^{2}}\left( \frac{\delta |\varepsilon _{0}|}{16\pi
c_{1}^{2}}\right) \frac{1}{\lambda _{ab}^{2}}.  \label{n2d}
\end{equation}%
By substituting (\ref{n2d}) into (\ref{critical1}) the critical temperature $%
T_{c}$\ takes the form
\begin{equation}
T_{c}=\left( \frac{\hbar c}{2\pi k_{B}e}\right) \frac{\left( 3\delta
|\varepsilon _{0}|\right) ^{1/2}}{\lambda _{ab}}.  \label{critical2}
\end{equation}%
Notice that $T_{c}$ does not depend on the pairon speed $c_{1}$. The
physical parameters such as $\omega _{D}$, $\varepsilon _{0}$, and $\lambda
_{ab}$ are suitable to be determined by a number of experimental techniques
\cite{fujita96,poole95}, while the layer width $\delta $ may be estimated
from band-structure calculations \cite{krakauer88}. The parameter $%
\varepsilon _{0}$ may be alternatively estimated from the BCS energy
gap at zero temperature $\Delta _{0}=\hbar \omega _{D}/\sinh
[1/N_{0}V_{0}]$, valid for arbitrary coupling. Combining this last
expression with (\ref{cooper}) shows that $\varepsilon _{0}=\hbar
\omega _{D}-\left[ (\hbar \omega _{D})^{2}+\Delta _{0}^{2}\right]
^{1/2}$ holds. Furthermore, in the weak-coupling limit, $\Delta
_{0}=\hbar \omega _{D}\exp [-1/N(0)V_{0}]$, and $|\varepsilon
_{0}|=\Delta _{0}^{2}/2\hbar \omega _{D}$. In that case, formula
(\ref{critical2}) may be conveniently rewritten and leads to the
simple formula for the critical condensation temperature%
\begin{equation}
T_{c}=\left( \frac{\hbar c}{2\pi k_{B}e}\right) \left( \frac{3\delta }{%
2\hbar \omega _{D}}\right) ^{1/2}\frac{\Delta _{0}}{\lambda _{ab}}.
\label{critical3}
\end{equation}%
This equation yields an alternative expression applicable in HTSCs\ for the
BCS ratio $2\Delta _{0}/k_{B}T_{c}\simeq 3.53$.

\begin{table}[b]
\caption{Physical parameters of cuprate superconductors and predicted
critical temperatures calculated according to formula (\protect\ref%
{critical2}). Parameters extracted from tables in Refs. \protect\cite%
{poole95,harshman92,hasegawa89,ginsberg92}. Debye temperature: $T_{D}\equiv
\hbar \protect\omega _{D}/k_{B}$. $\protect\delta=0.67 d$, where $d$ is the
CuO$_2$ layer separation in a given cuprate. }
\begin{center}
%\begin{ruledtabular}
%\par
\begin{tabular}{|ccccccc|}
\hline
& $T_D$ (K) \hspace{1mm} & $\Delta_0$ (meV)\hspace{1mm} & $\lambda_{ab}$
(nm) \hspace{1mm} & $\delta ( \mathring{A}) $ \hspace{1mm} & $T_c^{exp} (K)%
\hspace{1mm}$ & $T_c^{th}$ (K) \hspace{1mm} \\ \hline
\hspace{1mm} (La$_{.925}$Sr$_{.075})_2$ CuO$_{4}$ \hspace{1mm} & 360 & 6.5 &
250 & 4.43 & 36 & 36.4 \\
YBa$_2$Cu$_3$O$_{6.60}$ & 410 & 15.0 & 240 & 2.15 & 59 & 56.0 \\
YBa$_2$Cu$_3$O$_{6.95}$ & 410 & 15.0 & 145 & 2.15 & 93.2 & 92.6 \\
Bi$_2$Sr$_2$CaCu$_2$O$_{8}$ & 250 & 16.0 & 250 & 2.24 & 80 & 72.2 \\
Bi$_2$Sr$_2$Ca$_2$Cu$_3$O$_{10}$ & 260 & 26.5 & 252 & 2.24 & 108 & 109.2 \\
Tl$_2$Ba$_2$Ca$_2$Cu$_2$O$_{8}$ & 260 & 22.0 & 221 & 2.14 & 110 & 104.1 \\
\hspace{3mm} Tl$_2$Ba$_2$Ca$_2$Cu$_3$O$_{10}$ \hspace{3mm} & 280 & 14.0 & 200
& 4.30 & 125 & 105.5 \\ \hline
\end{tabular}
%\end{ruledtabular}8
\end{center}
\end{table}

According to (\ref{critical3}), for fixed values of $\omega _{D}$, $\Delta
_{0}$, and $\delta $, the critical temperature should increase linearly with
$\lambda _{ab}^{-1}$. This dependence has been indeed observed in
experimental studies of the correlation between the superfluid density, $%
n_{s}$, with critical temperatures of severely underdoped YBCO crystals,
with $T_{c}$s ranging from 3 to 17 K. The doping $p$ is the number of holes
of copper atoms per CuO$_{2}$ plane. Broun $et$ $al.$ \cite{broun07} found
that their samples of high-purity single YBCO crystals followed the rule $%
T_{c}\propto \lambda _{ab}^{-1}\propto n_{s}^{1/2}\propto (p-p_{c})^{1/2}$,
where $p_{c}$ is the minimal doping for the onset of superconductivity. A
similar behavior has been observed by Zuev $et$ $al.$ in YBCO films with $%
T_{c}$'s from 6 to 50 K \cite{zuev05}. They reported that, within some
noise, all their data fall on the same curve $n_{s}\propto \lambda
_{ab}^{-2}\propto T_{c}^{2.3\pm 0.4}$, irrespective of annealing procedure,
oxygen content, etc. Thus, by assuming that, except for $\lambda _{ab}$ the
YBCO parameters are approximately independent of $p$, we introduce in (\ref%
{critical3}) the values: $T_{D}=410$ K \cite{poole95}, $\Delta _{0}=14.5$
meV \cite{poole95}, and $\delta =2.15\ \mathring{A}$ \cite{krakauer88}, to
get the relation $T_{c}=16.79/\lambda _{ab}$ ($\mu m^{-1}$K). In Figure 1,
adapted from \cite{zuev05}, we compare this latter relation with the
experimental data obtained for underdoped YBCO films, as well as data
pertaining to a higher doping regime. We observe that the theoretical curve
provides an excellent fit to the experimental measurements. On the other
hand, the measured value of the penetration length in underdoped YBCO
crystals systems is an order of magnitude bigger than in thin films \cite%
{broun07}, so that the specific values of the critical temperatures derived
from (\ref{critical3}) are not in such a good agreement as in the YBCO
films. It has been pointed out that YBCO films seem to behave more similarly
to other cuprates, like BiSrCaCuO or LaSrCuO, than YBCO crystals \cite%
{zuev05}.\newline

The theoretical values of $T_{c}$ for cuprates with different compositions
have been also calculated either using formula (\ref{critical2}). Here we
focus on the following cuprates: 2(La$_{.925}$Sr$_{.075}$)CuO$_{4}$, YBa$_{2}
$Cu$_{3}$O$_{6.60}$, YBa$_{2}$Cu$_{3}$O$_{6.95}$, Tl$_{2}$Ba$_{2}$Ca$_{2}$Cu$%
_{2}$O$_{8}$, Tl$_{2}$Ba$_{2}$Ca$_{2}$Cu$_{3}$O$_{10}$, Bi$_{2}$Sr$_{2}$Ca$%
_{2}$Cu$_{3}$O$_{10}$, and Bi$_{2}$Sr$_{2}$CaCu$_{2}$O$_{8}$. The
characteristic parameters for these materials have been taken from
tables reported in
Refs.\cite{poole95,harshman92,hasegawa89,ginsberg92}. Concerning the
layer of width $\delta$, for which no direct experimental
determinations are available, an upper bound for $\delta$ is found
by considering that $\delta <d$, where $d$ is the separation between
adjacent CuO$_{2}$ planes \cite{harshman92}. In fact, as mentioned
before, band-structure calculations for YBCO systems yield $\delta
\approx 2.15\ \mathring{A}$ or, since $d=3.25\mathring{A}$, then
$\delta \approx 0.67\ d$. We do not expect that this relation should
be radically altered for other layered cuprates, so that in the
evaluation of $T_{c}$ we have assumed that it is approximately valid
for other cuprates besides YBCO. In Table I we present the results
obtained using the above assumptions, together with the physical
parameters involved in the calculation. In most cases we find a
rather satisfactory agreement between the predicted and
measured values of $T_{c}$. We observe that, in general, the exact formula (%
\ref{critical2}) leads to lower $T_{c}$'s than the approximate one (\ref%
{critical3}), although the values differ at most by 10$\%$. We have not
attempted to provide an estimation of the uncertainties of our theoretical
results, since the accumulated data of the physical parameters involved in
the calculation show a wide scatter.\newline

In conclusion, HTSC can investigated using a BCS-like theory for a quasi-2D
BEC of Cooper pairs satisfying a linear dispersion relation in their total
or CM momentum. Simple expressions ensue for the critical transition
temperature $T_{c}$. The formulas derived provide a functional relation $%
T_{c}\propto 1/\lambda_{ab}$. Although this apparently disagrees with
Uemura's phenomenological relation $T_{c}\propto n_{s}$ \cite%
{uemura04,homes04}, Zuev $et$ $al.$ \cite{zuev05} have pointed out that most
data in the Uemura plot refer to cuprate samples which are not severely
underdoped.

We show elsewhere that all relevant 2D expressions derived in this work
arise as the limit $k_{B}T\delta /\hbar c_{1}\rightarrow 0$ of a more
general 3D BCS-BEC theory for layered materials, and that conventional 3D
results are recovered in the limit $k_{B}T\delta /\hbar c_{1}\rightarrow
\infty $.

The authors would like to thank M. Fortes and M. A. Sol\'{\i}s for fruitful
discussions. MdeLl thanks UNAM-DGAPA-PAPIIT (Mexico) IN106908 for partial
support and is grateful to the University of Connecticut for its hospitality
while on sabbatical leave.

\end{document}